\RequirePackage{fix-cm}
\documentclass[smallextended]{svjour3}       
\smartqed  
\usepackage{graphicx}
\usepackage{url}
\usepackage{textcomp}
\journalname{Experimental Astronomy}
\begin{document}

\title{Information Systems Playground - The Target Infrastructure
}
\subtitle{Scaling Astro-WISE into the Petabyte range}

\titlerunning{Information Systems Playground}        

\author{A.N. Belikov, F. Dijkstra, J.A. Gankema, J.B.A.N. van Hoof, R. Koopman}

\authorrunning{A.N. Belikov et al.}


\institute{A.N.Belikov \at
              Kapteyn Astronomical Institute \\
              University of Groningen \\
              Landleven 12, 9747AB \\
              Groningen \\
              The Netherlands \\
              \and
              F. Dijkstra, J.A. Gankema \at
              Donald Smits Center for Information Technology \\
              University of  Groningen \\
              Nettelbosje 1, 9747AJ \\
              Groningen \\
              The Netherlands \\
              \and
              J.B.A.N. van Hoof, R. Koopman \at
              IBM Nederland\\
              Postbus 9999 \\
              1006 CE Amsterdam \\
              The Netherlands 
              }

\date{Received: date / Accepted: date}

\maketitle

\begin{abstract}

The Target infrastructure has been specially built as a storage and compute infrastructure for
the information systems derived from Astro-WISE. This infrastructure will be used by several applications 
that collaborate in the area of information systems within the Target project.
It currently consists of 10 PB of storage and thousands of computational cores. 
The infrastructure has been constructed based on the requirements
of the applications. The storage is controlled by the Global Parallel File System of IBM. This file system
takes care of the required flexibility by combining storage hardware with different characteristics
into a single file system. It is also very scalable, which allows the system to be extended into the future, 
while replacing old hardware with new technology.

\end{abstract}

\section{Introduction}
\label{sec:intro}

An information system consists of many components, the actual composition can vary due to the specific features requested by the 
users of the system.
Nevertheless a core part of any information system is the data storage. In the case of a scientific information system 
the second most important component becomes the processing facilities used to analyse the data. In this paper we will 
describe the infrastructure created for the support of Astro-WISE and many other information systems based on Astro-WISE. 

This infrastructure has been created in the context of the Target project\footnote{http://www.rug.nl/target}. 
The Target expertise centre is building a sustainable economic cluster of intelligent sensor network information systems. 
It is aimed at the data management for very large amounts of data. Several prominent research groups and innovative 
businesses are working together to develop and improve this infrastructure suitable for the complex and scalable data systems.

Target is a collaboration between research groups from the University of Groningen, including OmegaCEN and the artificial 
intelligence group, research groups from the University Medical Center in Groningen and companies like IBM, Oracle, Heeii and Nspyre. 
The infrastructure is for a large part hosted by the Donald Smits Center for Information Technology (CIT)
of the University of Groningen. In order to put the resulting technology to use elsewhere, valorisation of the Target 
knowledge and technology is also very important. This effort is coordinated by the Target holding.

The Target project is an umbrella for many applications, which have some common features. First, all these 
applications need huge data storage, second,  they also require massive data processing, and finally, they all can separate the 
information they use in a data and a metadata part. The latter is essential for the development of the infrastructure 
as each project will need two types of storage devices: one to store the bulk of the data as files, and another to store 
metadata describing these data items in a database. 

Astro-WISE is the first information system which implemented this approach (see, for example,\cite{Johnson1}). 
This approach has also been followed by the 
information systems for the LOFAR Long-Term Archive, Monk, LifeLines, etc., in which Astro-WISE is or will be one of the 
key technologies. These applications are described in the next section.

For Target a large storage and processing infrastructure has been set up, of which the storage part is based on IBM GPFS. The
infrastructure has recently been extended to a capacity of 10 petabyte of storage and can be extended further if required.
GPFS combines several storage pools, using different storage hardware, into a common global file system. 
The infrastructure is also coupled to a number of processing nodes, and also to the compute cluster of the university of 
Groningen (3220 cores), and the soon to be installed Grid cluster.

In the following sections we will first describe a few of the applications, including the requirements they have. After that
we will give an overview of some other solutions we have either looked at or that are currently gaining a lot of interest. 
In the next two sections an overview of the solution used is given, followed by the first feedback from applications 
run on the new infrastructure.

\section{Target Platform Requirements and use cases}
\label{sec:requirements}

We will start with the description of a number of scientific information systems, which are currently using the
Target platform. The ``first-born'' system of these is Astro-WISE which has relatively moderate requirements with respect 
to the data storage (not exceeding 0.5 PB of data) compared for example with the LOFAR Long-Term Archive with 10 PB for 5 years. 
Following are the Artificial Intelligence group information system Monk, and LifeLines which concentrates mostly on the data processing.
Requirements from these systems and applications formed the pool of requirements and defined the 
composition of the Target infrastructure.

\subsection{KIDS/VIKING}
\label{ssec:KIDS}

Astro-WISE hosts the data and the data processing of the KIlo Degree Survey (KIDS) and the data of the VIKING survey. Both are 
public ESO surveys which are performed on telescopes with wide-field cameras. Astro-WISE was developed for 
the data processing of KIDS. 

KIDS will combine an observation of 1500 square degrees in 4 bands and will be 2.5 magnitudes deeper than the Sloan Digital Sky Survey
(SDSS). The survey will allow to detect weak gravitational lensing, to study dark matter halos and dark energy, to detect 
high redshift quasars and to search for galaxy clusters. All these scientific use-cases will require an intensive data processing.

The source of all data processing are the raw data from the Omegacam camera. The input data volume is defined by 
the outline of the CCD camera, i.e., each observation consists of 32 files. Each file is an array of 8,388,608 
floating point values. Due to compression the typical file does not exceed 8 MB. Nevertheless, to build a single image the system 
should combine all 32 files and use calibration files as well, which makes a final image of a 
one GB size. 

The KIDS survey will cover 20,000 square degrees, creating a data flow of 30 TB per year at least. Despite the raw data rate 
being relatively small, the data processing and calibration will multiply this value, so that the data storage for 5 
years will reach half a petabyte.

The VIKING survey is an infra-red complement to KIDS mapping the same area as KIDS in 5 bands (Y,Z,J,H and K). The survey will be 1.5 mag
deeper than the UKIDSS Large Area Survey. The data load for VIKING exceeds approximately 3 times the data load for KIDS 
(7.5 TB per month  vs 2.5 TB per month for raw and calibration data),  but Astro-WISE will handle only already processed
images of VIKING. Nevertheless, this will add data volume to store comparable with the KIDS data.

Along with KIDS and VIKING data, Astro-WISE hosts data for a number of other instruments, which increases the requirements 
to the storage space. All these data are necessary for the calibration of KIDS and in support of scientific use cases for 
the KIDS-related research.

\subsection{LOFAR Long Term Archive}
\label{ssec:LOFAR_LTA}

One of the biggest users of the Target infrastructure is the LOFAR long term archive (LTA). 
LOFAR\footnote{http://www.lofar.org}, the Low Frequency Array is a huge radio interferometer. It consists of a compact 
core area (approx. 3 km in diameter, 18 stations) as well as 18 remote Dutch stations and 8 
International stations. Each station is equipped with up to 96 high band antennas and 96
low band antennas. Each station is connected to the Central
Processing System (CEP) through a wide area network using 10
Gbit/s Ethernet. The LOFAR instrument performs an aperture synthesis at the
Central Processing System using signals from each station. CEP consists of
a supercomputer of the type IBM BlueGene/P, and a Linux based cluster
at the University of Groningen.
More details on LOFAR can be found in papers by Gunst and Bentum\cite{Gunst} and De Vos et al.\cite{DeVos}.

After first processing on the LOFAR CEP systems (located in CIT, Groningen), data is archived in the LOFAR Long Term Archive (LTA).
This LTA is distributed over a number of sites, currently including the Target facilities in 
Groningen (Netherlands), SARA\footnote{http://www.sara.nl} Amsterdam (Netherlands) and 
Forschungszentrum J\"{u}lich\footnote{http://www.fz-juelich.de} (Germany).

The LOFAR Information System will deal with the data stored in the Long Term Archive. 
This Information System has been described in two previous papers \cite{Valentijn,Begeman1}.

The LOFAR LTA will store several petabytes of data on the Target storage facilities. Most of this data 
will need further processing. For the LOFAR application the following requirements can be given:
  \begin{itemize}
  \item Large data sets; The LOFAR data sets can be up to several TB in size. Each data set consists
        of several sub-bands. These sub-bands can be processed on different CPU cores, although some global 
        steps are necessary.
  \item Multiple petabytes of data in total; LOFAR expects to store more than a PB of 
        data each year into the long term archive.
  \item Processing of data is required; The total processing requirements currently exceed the
        capabilities of LOFAR's own CEP infrastructure. Furthermore any reprocessing of archived
        data needs to be done within the archive itself. 
  \item Individual files are not continuously processed; Not all LOFAR data is continuously being
        (re)processed. This data can therefore be moved to less expensive and slower storage.
  \end{itemize}

\subsection{Other applications}

Within the Target context there are several other non-astronomical applications that will make 
use of the Target infrastructure. Since their requirements also have been taken into account 
we will give a short description of these applications.

The first application worth mentioning is the recognition of handwritten text. The Artificial 
Intelligence group of the University of Groningen is working on this topic\cite{Zant1,Schomaker,Zant2}. 
For this application, called Monk\footnote{http://www.ai.rug.nl/~lambert/Monk-collections-english.html}, a massive amount of page scans of books is stored and
processed. This processing has two phases. The first is splitting up the pages into lines and
word zones. This step involves a lot of image processing techniques. The second step is identifying
words in the scans using intelligent matching algorithms and a self learning system. This system takes
input from volunteers who are labelling words. After this step these users are presented with words the system thinks are the same. 
The corrections applied by the volunteers are used to make the matches better. 
The whole system is based on searching in the scanned handwritten text and does not depend on conversion from the handwritten 
text into an ascii form.

The requirements that come from this application are, besides the data storage, the possibility to deal with 
many small files and the possibility to perform the preprocessing and learning phases on attached compute nodes.

A second application comes from bioinformatics. The LifeLines study will construct a bio-databank, populated with data from approximately 165,000 people. The need for accessibility, the size and the huge diversity in types of data (images, graphs, numbers, including genome and phenotype data) that need to be included make this a challenge. Moreover, at present it is still unknown which data will eventually be queried, and therefore, broad data mining needs to be facilitated. The Target approach as used in Astro-WISE is therefore eminently applicable. Requirements from this area are besides the data volume and number of files, also security and privacy issues. 

Within Target collaborations with commercial parties are also being started. This is part of the valorisation
of the Target infrastructure and knowledge. Because projects are currently in a start-up phase, and third parties are involved we can  not go into detail here.

\subsection{Databases}

Because the mentioned applications make use of the information technology developed in Astro-WISE, storage for their databases is 
also required. In principle having a fast storage platform for database access is an advantage. Furthermore the data in the database 
is of crucial importance, both in terms of reliability as in availability. Good performance for database applications and a 
high level of redundancy and availability are therefore also requirements.

The fraction of the metadata to data varies for different
applications but usually metadata does not exceed 10\% of the data volume. In the case 
of Astro-WISE only 3\% of the data volume is metadata stored in a relational DBMS, and the rest is files. Other use cases have an 
even higher data to metadata ratio. 
Currently Astro-WISE, the LOFAR Long-Term Archive and other projects are using an Oracle 11g RAC installation to store the metadata. 
An Oracle Golden Gate solution is used to mirror and replicate metadata databases to other, non-Target nodes.

\subsection{Pool of requirements}

The use cases described above differ in the way they use resources or organize the data, but they have something in common as well: 
all of them have a clear separation between the data (files with the images from a CCD, LOFAR telescope data, scans of book pages) and 
metadata (description of these data files). Moreover, in all these cases the main data volume is in the data.  

There are a number of problems for the data storage which the Target platform should answer: 
\begin{itemize}
\item The data volume, which will exceed 10 PB within the next 5 years.
\item The differences in the data itself. Target hosts a number of information systems. Astro-WISE and LOFAR deal with ten thousands of 
files of GB and TB size. For other applications like Artificial Intelligence one must store millions of files of MB-size.
\item A platform for the databases must also be provided. For this storage good performance, high availability and redundancy are requirements.
\item Authorization and administration problems. All the data stored on the Target platform should be available to the authorised user of the data only and protected from unauthorised access. It should be possible to combine the proposed
authorisation mechanism with the already existing A\&A systems for Astro-WISE and LOFAR (see Section 3.4 of~\cite{Begeman1} for details).
\item On-line and near-online storage. The access frequencies of different files differ; Some of the LOFAR files will be accessed once-per-month
or even once-per-half-a-year, meanwhile other files of Astro-WISE and the LOFAR Long Term Archive will be accessed daily. The access pattern will also change over time. To make the solution cheaper it should combine expensive fast disks with much cheaper tapes. The solution should provide a mechanism to bring files
from near-online to online storage using an analysis of the request rate to these files.
\item Finally, an important requirement is the ability to run existing applications without significant changes in the code.
\end{itemize}

Some of the requirements described above are opposite to each other. On the other hand the solution should be a single system with 
different types of storage, not a set of different storage platforms. Due to the huge number of files, the size of some of the data sets, 
and the changing access patterns over time we have chosen for the Global Parallel File System (GPFS)\cite{Schmuck} by IBM as 
the solution for the file system layer for the storage infrastructure for Target.

\section{Investigated Solutions}
\label{sec:solutions}

The chosen storage solution for Target is based on IBM GPFS. It is of course very useful to compare this GPFS solution 
with other possible solutions that are available for solving the data storage and processing problems that the 
Target environment is trying to solve. More details on the GPFS solution itself can be found in the next section.

When looking at other storage solutions it is useful to first look at the requirements we had. These
requirements are mainly derived from the application requirements from the previous section. When combined with a number
of operational requirements we arrive at the following list of requirements:

\begin{itemize}
\item Proven scalability to petabyte range; The solution chosen should be in use on a similar
      scale at other sites.
\item The long lifetime of an archive means that the system should be extendable and that hardware will be replaced.
      These operations should have minimal impact on operations.
\item Support available; The solution should be supported by a vendor or large community.
\item Tape storage possible, including integrated lifecycle management; Given the petabyte scale the use of 
      tape storage can be very cost effective. In order to manage the disks and tapes the use
      of information lifecycle management is necessary.
\item Hardware and software sold independently; Given the required lifetime of the infrastructure of
      at least more than a decade, and the European rules for procurement for public bodies a separation between
      software and hardware acquisitions is very important.
\item Support for POSIX I/O; The need for running existing applications requires support for the POSIX I/O standard.
\item Parallel file system; The multi petabyte scale of the archive and the requirement for processing of
      data sets on the order of terabytes requires fast I/O to large chunks of data. Parallel file systems are
      a huge benefit in this case.
\item Mirrored setup; For data for which high availability is important a mirrored setup is essential.
\item Wildly different access characteristics. For some applications like databases a high number of I/O operations
      per second is important. For other applications a large streaming bandwidth is important. Some applications
      work with millions of files, others with very large files. Furthermore the access pattern changes over time.
\end{itemize}

Although most of these requirements relate to in-house solutions based on traditional file systems, it is useful to also look at alternative 
technologies like Cloud storage, Mapreduce and the related storage technology (like deployed by Google and Yahoo!) and Grid storage solutions.

\subsection{Parallel file systems}

After taking the above mentioned requirements into account the two remaining interesting options we considered when studying parallel 
file systems at the end of 2008 were GPFS\cite{Schmuck} (version 3.2.1) and Lustre\cite{Lustre} (version 1.6.6). Both of these were 
operating at a petabyte scale at supercomputer centres around the world, and hardware and software are independent from each other.

Both solutions are also able to meet the bandwidth requirements of applications working with large data sets. Both systems can also 
be extended quite easily. When we compared the solutions the main differences we found were in the support for redundant storage and 
near-line tape storage. GPFS supports both mirrored storage, where all blocks are stored twice, and has integrated lifecycle management 
capabilities, where data can be moved transparently between different storage solutions, including tape storage.

Therefore the main reasons for selecting GPFS over Lustre were the information lifecycle management capabilities included with GPFS, 
the straightforward integration of tape storage, and the possibilities of creating more redundancy using mirrored setups. Furthermore a 
collaboration with IBM was started which would also give benefits in the area of support and future developments.

\subsection{The Cloud}
\label{sssec:clouds}

An important development in recent years is the rise of Cloud computing. Here services
are taken from the "Cloud", without the user worrying about the details of the specific
infrastructure. Important keywords are 'on demand' - you just use what you need, 
'scalability' and 'virtualisation'. Cloud computing ranges from infrastructure as a service, 
software as a service and platform as a service. For a good discussion on cloud computing 
we refer to a study performed by the University of California Berkeley\cite{Armbrust}. 
Note also that Cloud computing is in many ways related to the older 
concept of Grid computing\cite{Foster}.

In principle one could argue that the Target infrastructure is a Cloud in itself. It offers 
services to the user, where the user does not have to worry about the details of the services 
and of the hardware beneath. The Target infrastructure is one of these underlying components. 
The scalability of the GPFS platform can be used to seamlessly expand the Target infrastructure
without affecting the users.

One of the technologies that is interesting to compare with when looking at Cloud offerings is
the Amazon cloud, consisting of EC2 computing and S3 storage. In principle the Target
applications could also make use of such an infrastructure. An interesting study in this
area has been performed by Berriman et al. \cite{Berriman1,Berriman2}. In this study a comparison is
made between the price and performance of the Amazon cloud and a local HPC cluster.

The interesting observations are the following:
For I/O intensive applications the performance of the local HPC cluster using a parallel file system (Lustre in their case) is much better. 
For other applications the performance of the HPC cluster and of "large" instances of Amazon EC2 does not differ that much 
from each other. The specific resource usage is very important for the actual costs of using Amazon EC2. This because several things, including compute cycles, storage space and network usage are being charged. Especially the storage and transfer costs can have a large impact.

According to their findings a local HPC cluster is cheaper than the use of Amazon EC2. Note that the data storage
and I/O is largely determining the costs for Amazon EC2. Furthermore they assume a rather constant 
workload over time for the local facility, because only the costs of a single compute node were taken into
account. For larger scale local facilities changing usage patterns over time
should not be a big problem though, and the costs will indeed average out.

Given the large data volumes necessary for applications like LOFAR, where I/O bandwidth is also crucial,
Cloud storage solutions don't seem to be really cost effective for Target. Smaller applications will immediately benefit from the large scale required by the large applications as well.

\subsection{Mapreduce approaches}
\label{ssec:mapreduce}

Another solution worth looking into is the MapReduce approach and the underlying 
storage solutions. The MapReduce framework has been first introduced by Dean and Ghemawat of Google\cite{Dean}. Basically the framework consists of two steps, a Map step where input data is processed in parallel, and a Reduce step where the data from the Map step is combined into the desired result. 

An important aspect in the MapReduce approach is the availability and the location of the data. For this Google created the Google file system\cite{Ghemawat}. Here data is stored on a large cluster of commodity computer systems, using internal disks. Data is stored in  large chunks and those chunks are replicated over a number of nodes to provide reliability. MapReduce processing jobs are scheduled close to the data, preferably on the nodes that have the data. This greatly reduces the amount of network traffic. 

The proprietary work done by Google inspired others to develop Hadoop\footnote{http://hadoop.apache.org/}. Hadoop has been used for processing astronomical data by Wiley et al.\cite{Wiley}. Unfortunately in the paper no results on the effectiveness of this solution are given. We can point out some important differences with the Target approach though. For this we will especially focus at the file system.

One of the main differences is that the Hadoop file system (HDFS) does not support POSIX I/O. This means that legacy applications need to be rewritten to be able to make use of the advantages Hadoop may offer. HDFS is also oriented towards the MapReduce approach. Not all Target applications can be mapped to this paradigm.

The other main difference is that HDFS is optimized for large files. Typical file sizes are in the order of Gigabytes to Terabytes. This again means that applications have to be adapted. Several Target applications are currently using many small files. Applications that do use large files make use of many small files too. 

Still it may be useful to look at these concepts and see how they can be used within the Target context. A study by IBM also showed that it is possible to run Hadoop on top of GPFS\cite{Ananthanarayanan}.

\subsection{Grid}
\label{ssec:grid}

The last technology we will look at is Grid\cite{Foster}. Grid technology combines computing and storage infrastructure within different administrative domains. The technology is especially useful for people working together in different organisations
who want to make use of combined resources. Astronomical collaborations are a good example of such groups. The Grid concept is already in use for the Astro-WISE environment. The Astro-WISE environment is also able to submit jobs to the European Grid infrastructure operated by EGI\footnote{EGI has taken over operations of the European Grid infrastructure from the EGEE project, see http://www.egi.org}\cite{Begeman2}. Another example is the LOFAR LTA, which combines resources from different parties. 

When focussing on the storage infrastructure we find that there are several Grid storage solutions that try to combine a Grid interface adhering to the SRM standard\cite{Sim} with internal storage management. Two examples in use within the EGI context\cite{Stewart}, that also make use of hierarchical storage, are dCache\footnote{http://www.dcache.org} and Castor\footnote{http://castor.web.cern.ch/castor/}. Both manage a set of disk pools connected to a tape storage backend. The systems support load balancing and can work on a variety of hardware. In recent versions dCache is also able to represent its data in the form of a NFS 4.1 (parallel NFS) file system. This feature is still in development, however.

One of the main difference between the Grid storage solutions and the Target storage solution is the support for POSIX I/O. Only recently NFS version 4.1 support has been added to dCache. 

The only SRM solution, in operation within EGI, that natively supports POSIX I/O by design is StoRM\footnote{http://storm.forge.cnaf.infn.it/}. StoRM is able to run on top of a regular file system. It also supports GPFS very well and could therefore be perfectly installed on top of the Target storage.

\section{Design considerations} 

In order to comprehend the design considerations for the defined and deployed solution, there has to be a basic understanding of the IBM GPFS Solution.

The IBM General Parallel File System (GPFS) is a high-performance shared disk file management solution that provides fast, reliable access to a common set of file data from two computers to hundreds of systems.

GPFS integrates into other environments by bringing together mixed server and storage components to provide a common view to enterprise file data. It provides online storage management, scalable access and integrated information lifecycle management tools capable of managing petabytes of data and billions of files.

For the Target project the proven GPFS file management infrastructure provides the foundation for optimizing the use of  computing resources. The following properties of GPFS play an important role in order to be able to fulfil the application requirements described in section \ref{sec:requirements}:

\begin{itemize}
\item GPFS optimizes storage utilization by centralizing management; It allows multiple systems and applications to share common pools of storage. This allows dynamic allocation of storage space and the ability to transparently administer the infrastructure without disrupting applications. This also supports a heterogeneous workload reading from and writing to the GPFS clustered file system.

\item GPFS supports multi-vendor storage and server hardware, enabling striping, replication, and snapshots across heterogeneous systems. In the first hardware setup of the Target project all server and storage technology is based on IBM technology. 

\item Highly available grid computing infrastructure; 
GPFS is an essential component to implementing a high throughput grid environment. High-availability features including data replication and automatic failure handling provide a solid base on
which to build a research analytics grid or digital media management system.

\item Scalable information lifecycle tools to manage growing data volumes
One can use storage pools and the highly scalable policy engine to introduce structure to unstructured data. GPFS can manage billions of files using the integrated SQL based policies for file placement and migration over time. The scalable policy engine can be run on one system or concurrently on all systems depending on processing needs.

\item GPFS takes enterprise file management beyond a single computer by providing scalable access from multiple systems to a single file system. GPFS interacts with applications in the same manner as a local file system but is designed to deliver much higher performance, scalability and failure recovery by allowing access to the data from multiple systems directly and in parallel.

\item High-performance data access is achieved by automatically spreading the data across multiple storage devices and the ability to read and write data in parallel. In addition, for high-bandwidth environments like digital media or high performance computing (HPC) environments, GPFS can read or write large blocks of data in a single operation minimizing the overhead of I/O operations.

\item For optimal reliability, GPFS can be configured to eliminate single points-of failure. Availability is further improved by automatic logging and data replication. Data can be mirrored within a site or across multiple locations. The file system can be configured to remain available automatically in the event of a disk or server failure.

\item Central administration; GPFS simplifies administration by providing control of the entire system from any node in the cluster. Administration functions are based on existing UNIX and Linux administrative file system commands.

Heterogeneous servers and storage systems can be added to and removed from a GPFS cluster while the file system remains online. When storage is added or removed the data can be dynamically re-balanced to maintain optimal performance.

\item GPFS supports file system snapshots, providing a space efficient point in time image of a file system at a specified time. This provides an online backup to protect from user errors or a frozen view from which to take a backup.

\item GPFS supports mixed clusters of AIX, Linux and Windows 2008 (64 bit) systems in a single cluster. In addition multiple versions of GPFS can exist within a single cluster supporting rolling upgrades and providing operational flexibility.
\end{itemize}

After now having discussed all the relevant features of GPFS we can start to describe how these have been 
put to use within the Target infrastructure.

\section{Target Platform Setup}
\label{sec:setup}

The Target Expertise centre is building a durable and economically viable cluster environment, aiming at the operational management of extremely large    quantities  of data.  This scalable platform has now been installed and implemented in Groningen.

This installation has been made possible through and with Target partner IBM Nederland BV. IBM has played an advising role which covered the deployment of several different “High Performance Computing” (HPC) solutions, solutions which each already have a proven track record at implementations all over the globe.

Based on the requirements that have been gathered in 2009, for the delivery of the initial setup of the Target infrastructure IBM has architected  a solution based on open standards based IBM Hardware and Software building blocks, which has been installed and integrated by highly skilled IBM specialists.

\subsection{Schematic Overview Target multi-tier environment}

\begin{figure*}
\includegraphics[width=1.00\textwidth]{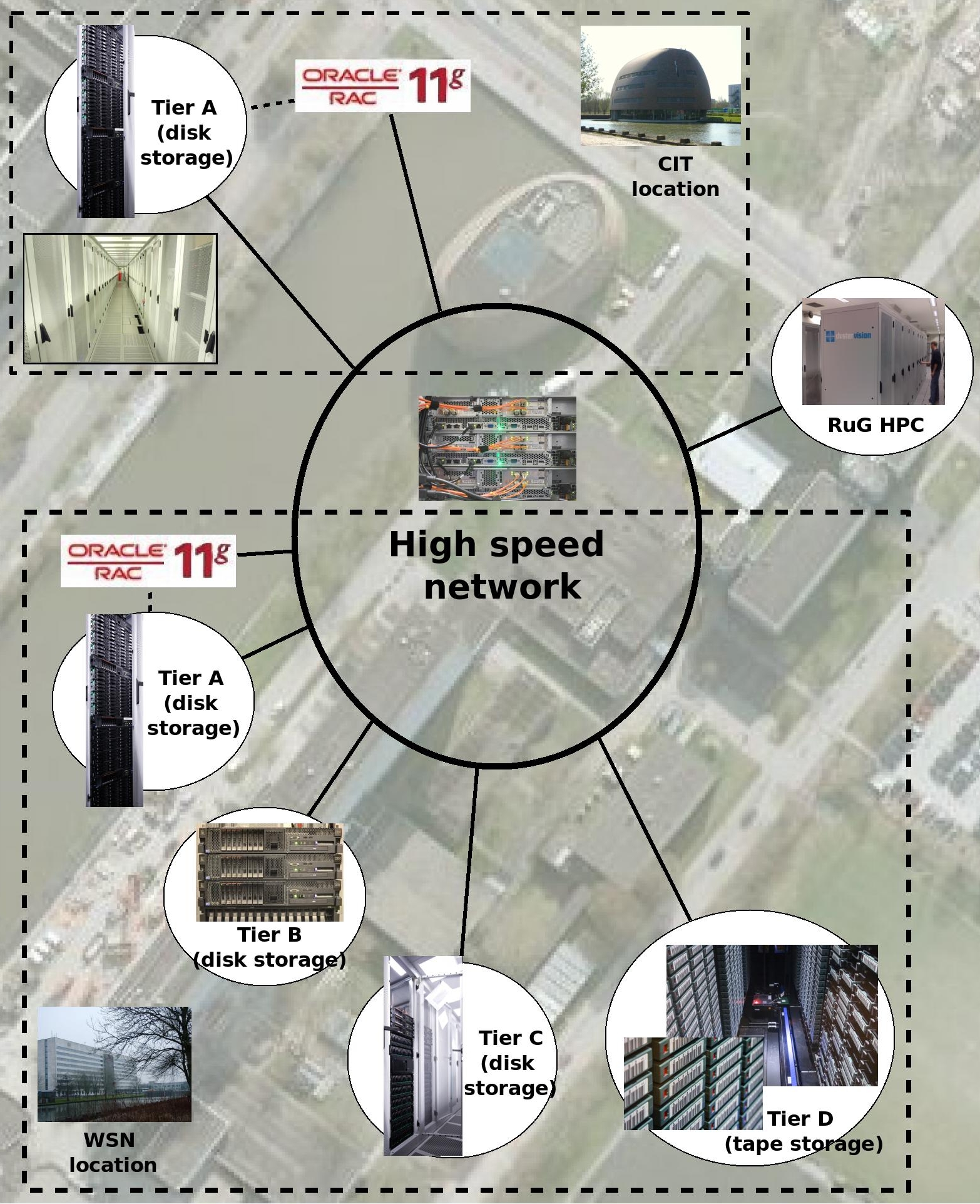}
\caption{Schematic representation of the Target platform}
\label{fig:target-schematic}
\end{figure*}

The Target infrastructure as originally envisioned and implemented by IBM can be very schematically represented as shown in Figure \ref{fig:target-schematic}.

This figure indicates the following properties of the solution:
\begin{itemize}
\item Multi-site - indicated by the two boxes in the figure. The infrastructure will be distributed over two data centres of 
the University of Groningen, located in different buildings. 
\item Multi-tier - multiple different kinds of storage solution have been implemented.
\item Distributed Database - as shown by Oracle RAC database servers being present at both sites.
\item High speed network - the network connects both locations and is a 40 gigabit ring with multiple connections for failover.
\end{itemize}
In the following sections properties of the platform are described. If needed a connection is made with the schematic overview.

\subsection{Information types}

In the Target infrastructure different pieces of information are of dissimilar types. The meta-data information will always be present in the database. The information that only encompasses data on a file system can be:
\begin{itemize}
\item reference information that needs to be available always;
\item measurement data (large size) that is analysed heavily the first weeks/months after acquisition;
\item derived data that is the result of the measurement data that exists in different versions.
\end{itemize}
The infrastructure can use the differentiation in types to optimize the cost of the solution used to store the information.

\subsection{Information importance}

The different types of information as mentioned before are of different importance for the users of the information. The importance of the information also changes over time. For example, derived information that has just been generated with a new type of model has a higher importance than that generated by an older model.

The importance associated with information is one of the indicators for the storage location that is to be used to store it. In general the importance a piece of information represents is determined by the user of the information, not by the IT organisation housing the information.

As information importance changes over time, the location of information can also change over time. The change of the importance and storage location of information over time is called the Life-cycle of the information.

\subsection{Information Life-cycle Management (ILM)}

Automated Information Life-cycle Management (ILM) is used to make sure information is always stored in the optimum location. This optimum can be characterized by many different parameters, the importance of the information being only one of these. ILM is one of the requirements for the Target infrastructure.

ILM is usually managed using policies set for different types of information or data. These policies are set by the owner of the information. The policy should follow the life-cycle of the information. So if information becomes lower in importance, 3 months after creation it could be moved to storage more suited to that importance level.
In general there are two types of policies:
\begin{itemize}
\item placement policy – this determines where information is stored on creation, i.e time=0
\item management policy – this determines the movement of information across different types of storage over time, i.e. time\textgreater0
\end{itemize}

The placement policy is activated once on creation of the information. The management policy is usually activated many times, according to a pre-determined schedule, e.g. once a day or once every hour.

A requirement to be able to perform ILM is to have different types of storage, representing different cost levels. This allows to optimise the costs for storing information related to an application. An infrastructure with these different types of storage is said to contain multiple tiers.

\subsection{Tiered storage structure / Storage pools}

A tiered storage structure consists of storage 'virtualisation' across multiple types of storage. The type of storage can differ along different dimensions, for example:
\begin{itemize}
\item size of the basic building block (small capacity HDD vs large capacity HDD);
\item performance of the basic building block (Fiber disks vs SATA disks);
\item energy consumption of the basic building block (HDD vs Tape);
\item cost of acquiring the basic building block (HDD vs SSD).
\end{itemize}

Typically all storage of a similar type is 'pooled' into a construct called a storage pool. In general a storage pool consists of storage with similar dimensions. The concept of different pools of storage enables the seamless integration of tape into the infrastructure.

As can be seen in Figure \ref{fig:target-schematic}, in the Target infrastructure 4 types of storage have been installed:
\begin{itemize}
\item tier A -- relatively inexpensive, Serial Attached SCSI (SAS) based harddisks, fast access, in a mirrored setup;
\item tier B -- large capacity, streaming, SATA based harddisks;
\item tier C -- medium capacity, high I/O per second, Fibre Channel based harddisks;
\item tier D -- large capacity tape.
\end{itemize}
In the Target infrastructure the tier A is mainly used for database files. Tier B can be used for data that can be streamed (e.g. larger files used in a sequential access pattern). Tier C enables a higher performance for random access of files. The tape tier, tier D, is essentially used of files that have an access pattern consisting of very infrequent access.

To ensure that some information stays available upon failure of part of the infrastructure, the storage is partitioned into failure groups.

\subsection{Failure groups}

A failure group is a collection of storage components that have a larger probability to fail at the same time. This can be storage components:
\begin{itemize}
\item housed in the same physical location – all fail at a site failure;
\item housed in the same rack – all fail when power to the rack fails;
\item connected to the same interconnect – all fail when the interconnect fail;
\item all running the same firmware – all fail when the firmware fails due to a bug.
\end{itemize}

Some of these failures are more probable than others.

For the Target infrastructure the tier A storage actually consists of two copies of everything. One copy located in a failure group in one data-centre and one copy located in a failure group in another data-centre. This is schematically represented in Figure \ref{fig:target-schematic} by the two boxes. The tier A storage contains the database files. This feature of failure groups corresponds to the requirement of a mirrored setup.

Given that storage components can be physically separated the networks and interconnects between them are important to maintain function and performance.

\subsection{Networks / Interconnects}

A large storage solution consists not only of storage components, but also of storage servers. In the case of the Target infrastructure database servers and application servers are also part of the solution. In this case several different networks or interconnects can be defined:
\begin{itemize}
\item storage interconnects -- connect storage components to storage servers;
\item cluster interconnect -- connects the storage servers, database servers and application servers at the level of the virtualisation layer (in this case a GPFS multi-cluster setup);
\item database interconnect -- connects the different database servers into a clustered solution;
\item management interconnect -- used to manage all components.
\end{itemize}

The storage interconnects can be split into several separate pieces. Some storage components are directly attached to servers using SAS connections, mainly in tier A. Other storage components, like tier B, C and D are connected to the storage servers using a FC SAN. The technology used for this is 8Gbit Fiber Channel.

The cluster and database interconnects are high bandwidth, low latency, Ethernet based networks. These networks run TCP/IP and enable communication between the server components. It is a 10 Gigabit Ethernet network.

The management interconnect is a TCP/IP based network as well. Since this does not need to be a high bandwidth network it is a Gigabit Ethernet network.

The different networks are schematically represented by a single ring in Figure \ref{fig:target-schematic}.

\subsection{GPFS handles the complexity}

The components described above create a complete solution that is quite complex. It is also a fact that the number and size of the different components will increase and decrease over time. If an end-user had to take all these characteristics into account, the Target infrastructure as used by the applications would be quite difficult to use. 

The Global Parallel File System (GPFS) solution of IBM takes care of all the complexity in the solution. It is a virtualisation solution on the file system level, which hides all the storage complexity from the end-user. The end-user 'sees' a POSIX compliant file system to be used by his applications.

GPFS also incorporates ILM functionality. The end-user only has to create rules for the importance of his information and how this changes over time. Essentially this means that he has to define the life-cycle of his information. These rules can then be translated into placement and management policies in GPFS. Data block movement is then automatic and invisible to the end-user. Even files that are in the process of being moved from one tier to another can be used and changed just like any other file.

\subsection{GPFS gives flexibility}
The GPFS solution enables the flexibility needed for the Target infrastructure. This flexibility consists of the following characteristics:

\subsubsection{Scalability}
Using the concept of storage pools and failure groups the GPFS solution becomes extremely scalable. Adding a building block of storage servers and storage hardware enables sizes of multiple petabytes. There is no single point in the GPFS architecture that becomes a bottleneck.

\subsubsection{Extendibility}
The building blocks in the GPFS solution can be added during operation. Adding a component can be done while the file system is online and in use. What is also important, as the Target infrastructure houses long term archives, is the fact that components can be removed using the same mechanisms. This means that older hardware can be replaced by first adding new hardware during normal operation and removing the old hardware afterwards, also during normal operation. GPFS will handle data block movement away from the old hardware and onto the new, modern hardware.

\subsubsection{Hardware agnostic solution}
GPFS software does not need specific IBM hardware to run on. Also the storage used to store the blocks for GPFS does not need to be of a specific type. This means that a GPFS based solution is quite hardware agnostic. This corresponds to the Target requirement about the fact that hardware and software should be sold independently.

\subsection{Underlying hardware solutions}
\label{ssec:hardware}

The underlying hardware installed for the first version of the Target infrastructure is as follows:

\begin{tabular}{lll}
{\bf Part }   & {\bf Hardware}                          & {\bf Raw capacity} \\ 
\hline\hline
tierA         & IBM DS3200, SAS disks, connected        & 56 TB \\
	      & through SAS links                       & \\
\hline
tierB         & IBM DS9900, SATA disks, connected       & 600 TB \\
              & through a storage area network (SAN)    & \\
\hline
tierC         & IBM DS5300, FC disks, connected         & 134.4 TB \\
              & through a storage area network (SAN)    & \\
\hline
tierD         & IBM TS3500, tape library, connected     & 900 TB \\
              & through a storage area network (SAN)    & LTO 4 \\
              & \& controlled by Tivoli Storage Manager & \\
\hline
Storage       & multiple IBM x3650 connected to the     & \\                                  
servers       & storage tiers, running SLES 11 \& GPFS  & \\ 
\hline
Database      & 4 IBM x3650, running SLES 11 \& GPFS      & \\
servers       & with Oracle RAC 11g ad DBMS               & \\
\hline
Application   & 4 IBM x3650                               & \\
servers       & running SLES 11 \& GPFS \\
\hline
Network       &  2 Brocade DCX 4 switches using 10GigE connections & \\
              &  40 Gbit/s interswitch bandwidth  &\\
\hline
SAN           &  2 Brocade DCX 4 switches using 8Gig FC connections & \\	     	           
\hline
\end{tabular}

\section{Application feedback}
\label{sec:feedback}

In this section we will describe some of the feedback obtained from the applications that have been using the newly installed Target infrastructure. This feedback describes the first migration steps and some first benchmarks run on the infrastructure.

\subsection{Integration with Astro-WISE}

The "dataserver" is native data storage middleware of Astro-WISE, which allows the Astro-WISE user to find any file 
in the system by its name. The Astro-WISE dataserver is a python-coded server which takes the name of the file as input parameter. It returns the file to the user if it is found on the managed file system, or the link to another dataserver if the file is not found locally. 
All dataservers are connected to build a mock-up of a global file system where the only option available to the users is to get the file just by using the name of the file. The file name is unique in the Astro-WISE system.

The use of the Astro-WISE dataserver on GPFS does not create any  problems as a dataserver can be installed on top
of any file system. Nevertheless the dataserver must not manage a file system ``too big'' as the search of the file in the underlying file system may take a long time in the case of GPFS due to the huge number of files. 
As a solution it is possible to install a cluster of dataservers which will communicate with each other and manage only a part of the GPFS file system. 

The problem with the dataserver as well as with any separation of the data storage and data processing is normally the necessity to transfer the file with the data from the storage facility to the data processing one. 
This necessity appears even in the case of the ``local'' data storage - the dataserver currently does not allow block level access to the file, but requires to transfer the whole file to some temporal storage on the user side.

A possible modification in the dataserver is to give to the ``local'' user on the Target platform not the URL to the file but a path on the local GPFS system, so that user can access the file from the application without copying it first. This option requires that the processing facilities, like a compute cluster, are connected to the data storage 
by installing GPFS clients on the processing side. 

A first setup where the Target storage is locally mounted has been made for the university of Groningen compute cluster Millipede. For the Grid cluster which is to be installed in the near future this requirement is part of the design.  

\subsection{Oracle on GPFS}

The Oracle database used for some of the applications like KIDS/VIKING and LOFAR is now running on top of the GPFS storage. This should give a benefit when performing full table scans. However, since the Astro-WISE software tries to avoid these, the switch to GPFS did not have much impact on the database performance right now. Since the database is expected to grow in the future the file system may have a larger impact in the future.

Another thing, which is currently of more importance, is the mirrored storage used for the database. This greatly improves the availability and reliability of the database storage, and is therefore extremely important for the availability of the Target applications.

\subsection{LifeLines application}

The first work that has been performed is the porting of the MOLGENIS software to the Target infrastructure. Another important application is the LifeLines research portal. In this portal the researchers can search for and work with the data on genotypes and phenotypes of all 165,000 LifeLines participants.

Another project is the "Genome of the Netherlands" project, in which the DNA of 750 Dutch people is completely sequenced and studied. For this task 100 TB of raw measurements have to be processed into 300 TB of resulting data. The data has been stored on the Target infrastructure and is processed on the University cluster, which has access to the data.

The collaboration between the different research groups and the computing centre within Target has been very important in this respect.

\subsection{Monk application}

\begin{figure*}
\includegraphics[width=1.00\textwidth]{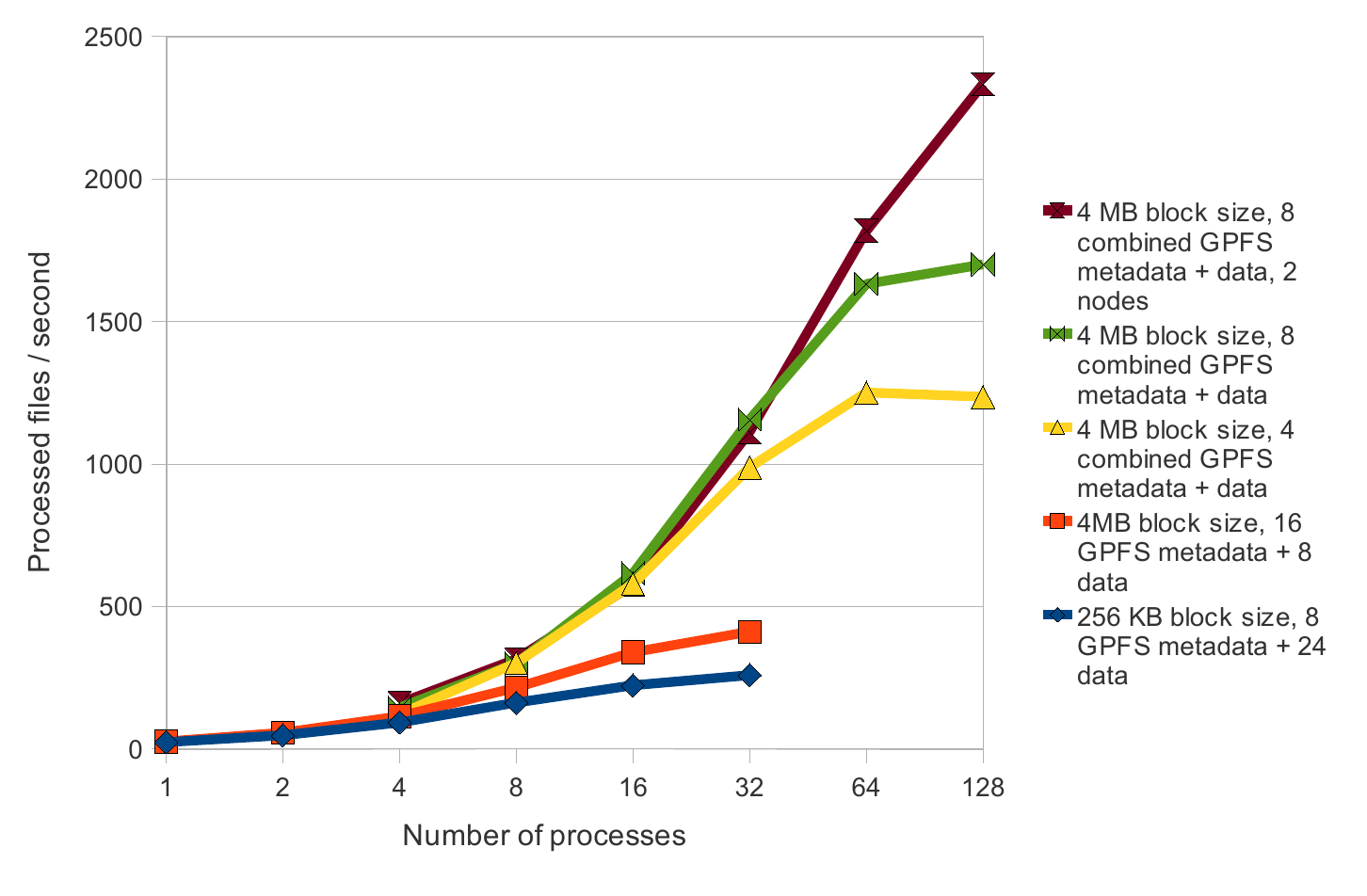}
\caption{Results of handwritten text processing benchmark on different GPFS file systems}
\label{fig:monk}
\end{figure*}

One of the first applications that has been tested on the Target infrastructure is related to the processing of handwritten
text. This processing involves splitting books and pages into many small parts, after which a learning process seemingly randomly
jumps through this files. This sort of processing is very demanding for storage system. Therefore a test application has been constructed, 
that mimics this random access on many files behaviour. The main problem is caused by the huge number of files distributed over a large number of directories. A nested directory tree is used, with one directory per book (for 50 books), per page (for 1500 pages), per line (for 30 lines). This with 30 files per line. Using the results of this application some tuning has been performed on the GPFS installation.
The results for the test are shown in Figure \ref{fig:monk}.

The figure shows results for a few GPFS file systems, created on the Target infrastructure. Without giving all the details the following conclusions
can be drawn:
\begin{itemize} 
\item The block size does not matter that much, as can be seen from the first two results in the graph. 
\item The number of file system metadata disks has a clear impact.
\item The different type of disk does not matter that much, since the fastest disks were used for the first two tests.
\item Combining file system metadata and data really helps.
\item The main problem is caused by the huge number of inodes as operated on by the application. 
\end{itemize}

The main result from this test is that for an application like this combining the file system metadata and data on the same disks works best. It would be
even better to try to adapt the application to make it less demanding in terms of the inode operations.

\section{Conclusion and future work}
\label{sec:conclusion}

In this paper we have described the Target infrastructure. This infrastructure will be the basis for several information systems 
derived from Astro-WISE. We have been able to meet all the relevant user requirements through the use of the Global Parallel File 
System (GPFS) by IBM. 
The infrastructure supports the storage of multiple petabytes of data in an efficient way. The use of several storage tiers helps 
in finding a good balance between performance and the costs related to the storage and the performance. The system is also very 
flexible with respect to user requirements and can be easily extended in the future. New hardware can be added and old outdated 
hardware can be removed, without impacting the user. 

The development of the Target infrastructure is crucial for the improvement of the performance of a number of systems. For example, 
in the case of Astro-WISE half of the processing time is spent on the Input/Output operations, including the data transfer from the 
storage to the processing node. The linking of the processing cluster of the CIT with the Target infrastructure via GPFS will allow to reduce this
latency and increase the processing speed for for example KIDS. Taking into account that Astro-WISE allows to reprocess all the data from the raw
image to the final catalogue, literally for each object requested by the user, the improvement in the processing speed is really handy.

Since the infrastructure has just been recently delivered there is still a lot of work to do in porting applications to the new infrastructure. Some of this work needs to be done in the following areas:
\begin{itemize}
\item Improving the Astro-WISE dataserver technology to be able to handle files which are locally accessible. 
\item Better definitions of the external interfaces that some of the applications require.
\item Extending the Astro-WISE security model to add extra checks at the file access level for applications that require this.
\end{itemize}
The exact definitions of the work need to be determined together with the applications.

Recently the infrastructure has already been extended. This has resulted in the following changes 
with respect to the setup described in section \ref{ssec:hardware}:
\begin{itemize}
\item Tier D has been extended to 6.1 PB using LTO5 technology
\item A new SATA Tier E has been added with 2.2 PB raw capacity 
\item The network has been extended to now have a 80 Gbit/s connection between the switches
\item The number of application nodes has increased to 20
\end{itemize}

This extension was necessary to be able to accommodate the bandwidth and storage requirements for some of the 
applications, especially the LOFAR long term archive. 

In our opinion the Target infrastructure is a showcase of an infrastructure for information systems that also shows how
the collaborating parties can benefit from both sharing knowledge and experience as well as resources, resulting
in a very flexible infrastructure which will be able to accommodate the specific needs of the different applications.

\begin{acknowledgements}
This work was performed as part of the Target project. This project is supported by 
Samenwerkingsverband Noord Nederland and Gemeente Groningen. It operates under the
auspices of Sensor Universe. It is also financially supported by
the European fund for Regional Development and the Dutch Ministry of Economic Affairs, Pieken in de Delta,
the Province of Groningen and the Province of Drenthe.
\end{acknowledgements}




\end{document}